\documentstyle[12pt]{article}

\newcommand{\beq}{\begin{equation}}
\newcommand{\eeq}{\end{equation}}
\def\reals{\hbox{\rm I\kern -.2em R}}
\def\com{\hbox{\rm l\kern -.4em C}}

\begin{document}
\begin{center}
\Large
{\bf TOWARDS QUANTIFYING NON-LOCAL INFORMATION TRANSFER: FINITE-BIT NON-LOCALITY\footnote{Paper presented at the Physics of Quantum Electronics Conference, Jan. 3-7, 1999}}
\normalsize
\vspace{.3in}

Michael Steiner \\
mjs@mike.nrl.navy.mil \\
February 3, 1999 \\
Naval Research Laboratory, Code 5340, Washington, DC 20375 \\
\end{center}

\vspace{.3in}

\begin{abstract} 
The advent of Bell's inequalities provoked the possibility that
entangled quantum phenomena is non-local in nature.  
Since teleportation only requires a finite amount of classical information, i.e. two bits,
the author asks whether or not it is possible to further characterize the nature of internal
correlation in terms of information.  Towards this end, the issue of the amount of information that
is transferred internally and non-locally is addressed. There are two possibilities: the amount
is infinite or the amount is finite.  A partial answer to this problem is 
given: it is shown that models exist whereby the amount is finite.
The EPR-Bell cosine correlation can be reproduced exactly using on average 1.48 bits.
The issue of simultaneity and the problems it poses are also examined in this context.  
Several extensions are suggested.
\end{abstract}

\vspace{.5in}

\noindent {\bf I. BACKGROUND}

\vspace{.2in}

Since the advent of Bell's inequalities and the subsequent verification of the
of the Quantum Mechanical predictions, it is speculated by some that there exists in nature
information transfer across large distances that occurs faster then the speed of light,
and perhaps, instantaneously. While such information transfer 
may occur internally to elementary particles, it is widely accepted that
internal information transfer cannot be utilized to develop a classical signaling system that 
transmits external information instantaneously.
However, that such internal information transfer may occur in nature is counter-intuitive.
The purpose of this paper is to develop a framework towards determining 
the amount of internal information (nonredundant) that is internally transferred non-locally. 

The author is motivated to investigate this for two reasons. 
Firstly, it is seen in the teleportation experiments that in addition to the 
Einstein-Podolsky-Rosen (EPR) correlation, 
only a finite amount of classical information is required, yet a particle with an
arbitrary density matrix is teleported. Hence, it is worthwhile to see if 
the teleported particle can actually be specified with only the initial state of the EPR particles,
and finite-bit non-locality (FBNL). 
Certainly, if only infinite-bit NL models could be proven to exist and no FBNL model exists, then the 
specification of a particle at a distance would require an infinite
amount of information, and the fact that teleportation is possible with only 
two classical bits would be astonishing from an information perspective.
If it can be proven that the amount of non-locality required
to replicate the teleportation experiments is greater then 2 bits, then teleportation 
must be utilizing non-locality (considered over the set of FBNL models). 
If the amount of non-locality required is less than or equal
to 2 bits, then a model may exist whereby the 2 bits that are classically transmitted contain the information
of the state of the particle, and the initial local hidden information is simply used as a reference.
Related questions have been asked by others, for example, Vaidman in \cite{Q1:Vaidman} questions if the essence of a 
state of a spin-1/2 particle is just 2 bits.

The second reason to investigate whether or not a FBNL model exists is 
that an infinite amount of information can only be transferred via a channel that is effectively noiseless.
That is, since the capacity of a typical channel behaves like $B \log(1+\mbox{SNR})$, where $B$ is the
bandwidth and SNR denotes the signal-to-noise ratio, the only
way such information could be exchanged is if the SNR is infinite ($B=\infty$ does not work).
Hence, if no FBNL models exists, then one is forced to accept that the internal channel is
effectively noiseless.  This is hard to accept, since no analogy exists classically.
On the other hand, no classical analogy exists for many quantum phenomenon,
hence this second reason is not as pervasive as the first.

In Section II, non-local information transfer will be formally defined. In Section III, we will
show several admissible models, including the finite-bit non-locality model.
It is known that models of non-locality such as Bohm's model as well as the models considered
in this paper have not been reconciled with special relatively and in particular, the issue
of simultaneity. In Section IV implications of this issue are examined

\vspace{.2in}

\noindent {\bf II. NON-LOCAL INTERNAL INFORMATION TRANSFER}

\vspace{.2in}

There have been numerous studies regarding quantum information.
However, the problem presented in this paper does not appear to have been 
solved (related results have recently been obtained independently by Brassard, Cleve, and Tapp \cite{Q1:Brassard} discussed
in Section V). 
In \cite{Q1:Aravind} the question of how many bits are required to 
transmit a qubit reliably is examined. The approach appears to be 
based on an approximation, and a fidelity or accuracy is defined.  
In \cite{Q1:Caves} several related information quantities are addressed.

In order to understand entanglement it is useful (in my opinion) to consider models. 
The class of models assumed here is that a positive amount of hidden information is transferred between particles.  The
information is hidden in the sense that it cannot be manipulated by external observers for
the purpose of sending superluminal signals.  In this paper, it is shown that a finite amount of information
can be used to quantify the amount of hidden information. In fact, a technique is shown whereby only
1.48 bits is required to replicate exactly the cosine correlation function expected for the EPR experiment (with spin measurement).

A model that provides the same expected values as quantum mechanics
for a given experiment will be termed an admissible model for the experiment.
The reader may note that the wavefunction in quantum mechanics
itself is an admissible model. However, it could be argued that the wavefunction
in quantum mechanics is only a partially complete model. 
Additionally, it is well known that the measurement problem, or wavefunction collapse
in the Copenhagen interpretation of quantum mechanics, does not have a well-defined model \cite{Q1:Hughes}.

The issue of how much, if any, information must be transferred between particles
in order to observe the correlations observed in the EPR experiment deserves attention.
There are three possibilities for a given reference frame: 
1) no information is transferred, 2) a finite amount is transferred, and
3) an infinite amount of information is transferred. 
A reference frame is assumed whereby the measurement devices are not moving and the
EPR source is fixed closer to one measurement device then the other. The problem of Lorentz invariance
and the issue of simultaneity is examined in Section IV. 

Although the first possibility above whereby no information is transferred is 
discussed at the end of Section III, the primary purpose of the 
paper is to investigate whether or not it is possible to model the particles and obtain the quantum mechanical correlation
with FBNL as opposed to infinite information models.

\vspace{.2in}

\noindent {\bf III. INTERNAL INFORMATION TRANSFER MODELS}

\vspace{.2in}

The experimentally shown violation of Bell's inequalities points to the possible existence of non-local
phenomenon in nature.  However, neither Bell's inequality nor quantum mechanics shed much
light on just how these non-local effects propagate.  Bohm's and other theories do provide some insight,
but they require that some non-local effects occur. For example, the quantum potential can 
change  instantaneously in Bohm's theory. However, it appears from
the equations that specify the trajectories for this experiment \cite{Q1:Bell}[p. 132]
that the trajectories are a function of the parameters of both Stern-Gerlach devices.   
Hence Bohm's theory does not shed light on how much non-locality is required.

We will examine the amount of information that is required to be transferred via models that allow for the non-local transfer of hidden information. Consider the EPR experiment shown in Fig. 1.  There is an initial state $\Lambda_{I}$ that characterizes the
initial state. Experimenter A has a Stern-Gerlach measurement apparatus oriented at $\theta_{a}$ which records
$A=1$ or -1 and experimenter B has a similar apparatus specified by $\theta_{b}$.  
The source is positioned closer to A then B. Two entangled spin 1/2 particles will be assumed to be emitted with
state $|\psi> = \frac{1}{\sqrt{2}}(|\uparrow>|\downarrow>-|\downarrow>|\uparrow>)$

We will now assume the possibility that classical hidden information
denoted by $\Lambda_{H}$ can be transferred through a channel with capacity $C_{H}$ bits/use, and that this channel is used
once so that only $C_{H}$ bits can be transferred.  
We assume that $\Lambda_{H}$ is chosen from a discrete
memoryless source so that its information content may be finite, given by $\mbox{H}(\lambda_{H})$. Note
that specifying a continuous value exactly requires infinite precision. In more precise terms, trying
to send a continuous source through a channel requires a capacity specified by the rate distortion
function $R(d)$ in information theory. As long as one allows for a given amount of distortion,
the information can be sent reliably.  However, if no distortion is allowed, i.e. $d=0$, then
for most continuous sources (such as the Gaussian source)  $R(d)$ is unbounded. 
Since we do not require to transmit the set $\Lambda_{I}$ at a distance, 
we assume that $\Lambda_{I}$ is a set of fixed complex numbers.  However, whether or not this assumption is required
will be examined in Section V.

Denote a finite-bit model (FBM) $M=\{\Lambda,R\}$ to be a set of variables $\Lambda={\lambda_{I},\lambda_{H}(\theta_{A})}$ 
and a measurement rule $R$= \{$A({\lambda_{I},\theta_{a}), B(\lambda_{I},\theta_{b},\lambda_{H}})\}$.
A model that agrees statistically with quantum mechanics for a given experiment ${\cal E}$ will be called an admissible model
for that experiment and is denoted by $M\in \cal{A(E)}$. Finite bit admissible models are only a subset
of $\cal{A(E)}$ and the admissible models for this class are denoted $\cal{A_{\mbox{FB}}(E)}$.  
The quantum mechanical predictions for this experiment are 
\beq \mbox{E}(A B) = -\cos(\theta_{a}-\theta_{b}) \eeq
with marginals $\mbox{E}(A)=0$ and similarly for $B$.

Now, there are two cases to be considered. The first
is where $C_{H}$ is an average constraint so that the amount of information can be more then $C_{H}$
on any one channel use. The second is a peak constraint so that for each channel use, only $C_{H}$ bits or less are
transmitted. Models $M$ for which there exists a source code characteristizing $\Lambda_{H}$ 
with lengths that are all less than or equal to $C_{H}$ bits are called peak constrained models 
and are denoted as ${\cal M}_{\mbox{peak}}$. 
Similarly models for which the source code has average lengths less than $C_{H}$ are denoted ${\cal M}_{\mbox{avg}}$.
In this paper, only the average constraint will be examined. See Section V for results regarding 
the peak constraint. Define
\beq C_{H,\mbox{avg}} = \min_{M\in\cal{A_{\mbox{FB}}(E)}, M\in {\cal M}_{\mbox{avg}}} \mbox{H}(\Lambda_{H}) .\eeq
and similarly for $C_{H,\mbox{peak}}$. Note that $C_{H,\mbox{avg}}\le C_{H,\mbox{peak}}$. 
We will now demonstrate an average constrained model for which $\mbox{H}(\lambda_{H})=1.48$ 
and hence $C_{H,\mbox{avg}}\le 1.48$  (for this experiment).

\vspace{.2in}

\noindent {\bf Generating an Arbitrary Variate at a Distance with Finite Information}

\vspace{.2in}

Consider first the following more general problem. Suppose that one would like to generate at B
a single random variate which has any distribution $f_{\theta_{a}}(x)$  that is parameterized by
a continuous parameter such as $\theta_{A}$.  
For example $f_{\theta_{a}}(x)$ could be a distribution with a peak at $x=a$.
Suppose furthermore that both A and B have available to them a random number seed specified by $\Lambda_{I}$, 
and some method of generating random numbers which is composed of variates from some {\it a prior} distribution. 
However, only A knows $\theta_{a}$.  Is it possible for A to send on average a finite amount 
of information to B so that B can construct
a single random variate with distribution exactly equal to $f_{\theta_{a}(x)}$? The answer is 
surprisingly yes, at least under certain conditions.  One solution is afforded by adopting the rejection method of non-uniform
random number generation\cite{Q1:Devroye}. The rejection method is as follows. Suppose that 
one can find a distribution $g(x)$ and a constant $c\ge 1$ so that \[ f_{\theta_{a}}(x) \le c g(x) \] for all $x$ and 
all $\theta_{a}$. Perform the following steps:

\noindent Step 1: Generate a variate $w$ with distribution $g(x)$ and another variate $u$ that is uniformly distributed on $[0,1]$. \\
Step 2: Compute the function $T=u~c~g(w)/f(w)$.  \\
Step 3: If $T>1$ repeat starting from step 1. If $T\le 1$ then $w$ has the desired distribution $f_{\theta_{a}}(x)$.  \\

Now, all that is needed for A to do is to send to B the iteration number where he first succeeded 
in getting $T\le 1$. B then simply generates this variate using the
same seed and has succeeded in generating a variate from the distribution $f_{\theta_{a}}$. One should note that the
amount of information required may or may not be finite since potentially the iteration at which
A succeeds is not bounded by any constant. Hence, the convergence of this process are significant. 
It is known that the convergence of this process is only a function of the parameter $c$, and is specified by the
geometric distribution, $P(K=k)=(1-p)^{k-1}p$, where $p=1/c$ and $K$ is the iteration number. In fact 
$P(K>k)=(1-p)^{k}\le \exp(-p k)$, hence the technique convergences exponentially fast.
From source coding theory, it is known that a discrete memoryless source can be coded by a variable
length code with average length equal (or nearly so) to its entropy. This is also the average number of bits that A must send
to B.   The entropy of this process is given by 

\[ \mbox{H} =\sum_{k} - P(K=k) \log P(K=k) \] which can be shown to be

\beq \label{eqnent} \mbox{H} = -\log p - \frac{(1-p)}{p}\log (1-p) .\eeq

Hence if one can find a $c\ge 1$ whereby $c g(x)$ that is uniformly larger then $f_{\theta_{a}}(x)$, then
a technique follows which only requires a finite amount of information.

\vspace{.2in}

\noindent {\bf Application to Bell pair}

\vspace{.2in}

This technique can be immediately applied to the Bell pair. It was shown in \cite{Q1:Feldmann} that if we
define $A(x,\theta_{a}) = \mbox{sign}(\cos(x-\theta_{a}))$, $B(x,\theta_{b}) = -\mbox{sign}(\cos(x-\theta_{b}))$
and $P(x,\theta_{a}) = 1/4 |\cos (x-\theta_{a})|$ then,
\beq -\cos(\theta_{a}-\theta_{b}) = \int A(x,\theta_{a}) B(x,\theta_{b}) P(x,\theta_{a}) dx \eeq
and $\mbox{E}(A)=\mbox{E}(B)=0$.  The rejection method above is applied at particle A to generate $f_{\theta_{a}}(x)=1/4 |\cos (x-\theta_{a})|$ using
a random number generator with seed specified by $\lambda_{I}$ and with g(x) uniformly distributed on $[0, 2\pi]$. 
We assume that a new seed is generated for every new experiment with a Bell pair.
Note that the condition $f_{\theta_{a}}(x)\le c g(x)$ is satisfied for $c=2\pi/4$ for all $\theta_{a}$.
Particle A measures $\mbox{sign}(\cos(x-\theta_{a}))$  and sends to $B$ the iteration number $k$ where A first
succeeded in $T\le 1$. Particle B then generates $x$ from seed $\lambda_{I}$ and can now directly generate 
$B(x,\theta_{b}) = \mbox{sign}(\cos(x-\theta_{b}))$ and therefore has succeeded in exactly generating the
cosine correlation function.  One can check that all marginals and the joint distribution are in agreement with the
Quantum Mechanical results as well. For this problem, the information required is specified by Eqn. (\ref{eqnent}) with
$p=1/c=\frac{4}{2 \pi}$. In this case Eqn. (\ref{eqnent}) reduces to $\mbox{H}=1.48$ bits. Hence for this experiment,
$C_{H,avg}\le 1.48 $ bits. A simulation was written and both the cosine correlation function and average
number of bits equal to 1.48 were verified.

\vspace{.2in}

\noindent {\bf Other models and a lower bound}

\vspace{.2in}

Note that the FB models considered have a particular form: $M=\{\Lambda,R\}$, $\Lambda={\lambda_{I},\lambda_{H}(\theta_{A})}$ 
and a measurement rule $R$= \{$A({\lambda_{I},\theta_{a}), B(\lambda_{I},\theta_{b},\lambda_{H}})\}$.
This constraint on the form  of the model means that we are considering only the subset
$\cal{A_{\mbox{FB}}(E)}\subset \cal{A(E)}$.
Note that upper bounds on a FB model class are still valid bounds if we enlarge
the class of models.  However, a non-zero lower
bound on $C_{(H,\cdot)}$ obtained for the class of FB models may not be valid if
we consider an enlarged model set. Hence $C_{(H,\cdot)}=0$ can not be ruled out very easily.
For example, a class of models could be envisioned that assumes that a deterministic function
exists so that all future events can be predicted locally given an initial state. If such a
model exists it is plausible that no non-locality is required to explain the 
EPR experiments. However, it is not clear that such a model exists.
If we restrict our model to the FB class, then it appears that a lower bound
should be attainable for at least this set.

\vspace{.2in}

\noindent {\bf IV. ISSUES OF LORENTZ INVARIANCE AND SIMULTANEITY}

\vspace{.2in}

It has been noted in several papers such as \cite{Q1:Hardy} among  others that realistic theories 
either require a preferred frame of reference or will give rise to well-known causal paradoxes. 
This has implications on causality when considering multi-simultaneity experiments \cite{Q1:Percival}. 
There are several types of causality violations that can emerge when considering such models. We will
consider two types. The first type of violation is that the experimenter can observe what he is about to do. The second violation is that the experimenter can not determine what he is about to do,  but what he is about to do 
must be some function of the future (but he can not determine what the function is).
This second violation is less serious then the first type, but bothersome nonetheless.  Interestingly, 
it turns out that the first type of violation can not occur with our model. This is due to the fact
that the distribution of the information that is sent from particle A to B is a geometric distribution which
is independent of the parameters of the detection devices, such as $\theta_{a}$.  
Furthermore, since the seed is hidden information and is non-accessible,
it is not possible for any external observers in any reference frames 
to glean anything about their future.  
However, it appears that multi-simultaneity experiments could be devised as in \cite{Q1:Percival} that would show the second type of violation occuring. 

Note that such problems are not limited to this theory. It is well known that other realistic models such
as Bohm's theory, have problems with Lorentz invariance. However, this problem still appears to be
significant . The relationship between entanglement and simultaneity has not been deeply explored. 
Some related work on this was done recently by Suarez \cite{Q1:Suarez}. This problem is also
related to collapse models. Both non-relativistic and relativistic wave function collapse models 
are reviewed in \cite{Q1:Pearle}. 

\vspace{.2in}

\noindent {\bf V. FURTHER WORK}

\vspace{.2in}

There is significant further work that can be done.  Note that $C_{H,avg}$ was not determined for general polarization
but for a particular polarization such as linear polarization.  It would be useful to generalize this result both for general spin 1/2 particles and general measurements.  In this paper, the EPR experiment with
two particles was analyzed.  However, it would be useful to analyze the teleportation problem which requires
three particles in more detail.  If it can be proven that the amount of non-locality required
to replicate the teleportation experiments is greater then 2 bits, then teleportation must be utilizing 
non-locality in addition to the 2 bits (considered over the set of FBNL models).  
If, on the other hand, the amount of non-locality required is less then or equal
to 2 bits, then a model may exist whereby the 2 bits that are classically transmitted contain the state of the particle (in
addition to the shared information $\lambda_{I}$).

The result that a finite amount of information can be used to obtain the cosine correlation function
was also obtained independently by Brassard, Cleve, and Tapp \cite{Q1:Brassard}. In fact, they
show that 4 bits can be used to obtain the cosine correlation function, hence have found
an upper bound on $C_{H,peak}$.  They also generalized their work
to arbitrary von Neumann measurements and showed that 8 bits would suffice.  This is very interesting, and their approach
is fairly different then the rejection method considered here. 
Their main interest is in Quantum Simulation, which is an application that the rejection method
could also be applied to. Hence we now know that $C_{H,peak}\le 4$ and $C_{H,avg}\le 1.48$ (for generation of the
cosine correlation function).

We have examined FB non-locality, but  with $\lambda_{I}$ chosen from the complex numbers.  Can this restriction
be removed so that $\lambda_{I}$ is chosen from a finite set with finite entropy? In this case, a spin 1/2 particle
could be completely specified with a finite amount of information.  Is $C_{H,peak}$ strictly larger then $C_{H,avg}$?  
Another generalization includes the considerations of higher-order spin, which requires all probability moments to agree.  
Also, can the teleportation of continuous quantum parameters as shown in \cite{Q1:Braunstein} be modeled by a finite-bit model.  
Does a general model emerge? 

\vspace{.2in}

\noindent {\bf ACKNOWLEDGEMENTS}

The author thanks Louis Sica for many valuable insights.  Thanks very much also for the suggestions from Richard Cleve and Alain Tapp, and others for discussions regarding this and future work.

\end{document}